\documentclass[prl,twocolumn,showpacs,preprintnumbers,amsmath,amssymb]{revtex4}
\usepackage{graphicx}
\usepackage{dcolumn}
\usepackage{bm}

\begin{document}

\title{Exploring  Interatomic Coulombic Decay by Free Electron Lasers}

\author{\firstname{Philipp~V.} \surname{Demekhin}}
\altaffiliation[Email: philipp.demekhin@pci.uni-heidelberg.de ]{}
\affiliation{Theoretische Chemie, Physikalisch-Chemisches
Institut, Universit\"{a}t  Heidelberg, Im Neuenheimer Feld 229,
D-69120 Heidelberg, Germany}

\author{\firstname{Spas D.} \surname{Stoychev}}
\affiliation{Theoretische Chemie, Physikalisch-Chemisches Institut, Universit\"{a}t Heidelberg,
Im Neuenheimer Feld 229, D-69120 Heidelberg, Germany}

\author{\firstname{Alexander I. } \surname{Kuleff}}
\affiliation{Theoretische Chemie, Physikalisch-Chemisches Institut, Universit\"{a}t Heidelberg,
Im Neuenheimer Feld 229, D-69120 Heidelberg, Germany}

\author{\firstname{Lorenz~S.} \surname{Cederbaum}}
\affiliation{Theoretische Chemie, Physikalisch-Chemisches Institut, Universit\"{a}t Heidelberg,
Im Neuenheimer Feld 229, D-69120 Heidelberg, Germany}
\date{\today}

\begin{abstract}
To exploit the high intensity of laser radiation, we propose to select frequencies at which single-photon absorption is of too low energy and two or more photons are needed to produce states of an atom that can undergo interatomic Coulombic decay (ICD) with its neighbors. For Ne dimer it is explicitly demonstrated that the proposed scheme to investigate interatomic processes by multiphoton absorption is much more efficient than with single-photon absorption of sufficiently large frequency as used until now. Extensive calculations on Ne dimer including all the involved nuclear dynamics and the losses by ionization of the participating states show how the low-energy ICD electrons and Ne$^+$ pairs are produced for different laser intensities and pulse durations. At higher intensities the production of Ne$^+$ pairs by successive ionization of the two atoms becomes competitive and the respective emitted electrons interfere coherently with the ICD electrons. It is also demonstrated that a measurement after a time delay can be used to determine the contribution of ICD even at high laser intensity.
\end{abstract}


\maketitle

Inner-valence (IV) ionized states of isolated atoms and small molecules are usually electronically stable and may only relax via fluorescence decay or dissociation. When embedded in environment the IV ionized system can efficiently relax on a femtosecond time scale by transferring its excess energy to its neighbors and ionizing them \cite{Cederbaum97prlicd}. In
the last decade, many theoretical and experimental studies of this process called Intermolecular or Interatomic Coulombic Decay (ICD) have been reported for various systems \cite{Averbukh,Hergenhahn}. The ultrafast relaxation driven by interatomic and intermolecular processes on the femtosecond time scale is of multidisciplinary importance in physics, chemistry, biochemistry, and biophysics  \cite{Averbukh,Hergenhahn,Marburger03,Jahnke04,Morishita06,Jahnke10,Mucke10,Aziz08,Kryzhevoi11,Stoychev11}.

There are many possibilities to initiate ICD in a system by single-photon absorption of sufficient energy to ionize an IV electron of the system. The new generation of light sources, free electron lasers (FELs), provide the opportunity to study radiation-matter interactions under extreme conditions, such as unprecedented high-intensity and ultrashort pulse durations. The presently operating FELs cover a wide range of short-wavelength radiation from VUV to soft X-ray (at FLASH facility \cite{FLASH}) and up to hard X-ray (at LCLS facility \cite{XFEL1}). Exposed to strong pulses, atomic and molecular systems may absorb a large number of photons creating differently charged ions and even bare nuclei \cite{XFEL2}. Moreover, if the energy of a single-photon is insufficient to produce a selected state, the high-photon flux of a laser pulse enables the production of such states by multi-photon absorption \cite{Saalmann06}. It is, therefore, rather natural to exploit these advantages of FELs for studying ICD processes.

Very recently \cite{Averbukh09,Kuleff10}, two possibilities to initiate ICD by multi-photon absorption in a cluster were discussed. In the collective ICD suggested in Ref.~\cite{Averbukh09} two atoms are IV ionized by two photons in a cluster where a single IV vacancy cannot trigger ICD. The simultaneous relaxation of the two IV vacancies can, however, undergo ICD by ionizing a third neighbor. In the process discussed in Ref.~\cite{Kuleff10}, two or more atoms are excited and the relaxation of one of the excited atoms results in the ionization of  one of the other excited atoms. As discussed in Refs.~\cite{Averbukh09,Kuleff10}, in intense fields these processes can provide efficient multi-photon ionization pathways in matter.

In the mechanisms discussed in Refs.~\cite{Averbukh09,Kuleff10} one cannot view the respective processes as the relaxation of an ionized or excited atom embedded in an environment as also a neighbor has to be ionized or excited. In the present work we intend to close the gap by investigating an atom which absorbs at least two photons and decays via ICD by ionizing its neighbor, providing a hitherto unexplored possibility to initiate ICD by an intense FEL pulse. Our results pave the way for further studies of interatomic processes by strong pulses of low-energy photons. A single absorbed photon which ionizes an outer-valence (OV) electron of an atom (or molecule) embedded in environment does not trigger ICD. On the other hand, irradiating the system by an intense laser pulse with energy above the OV, but below the IV ionization threshold, can induce ICD. This particularly applies if the central frequency of the pulse is also tuned to the energy difference between the IV and OV ionized states of the atom as these states will be strongly coupled by the intense field. For isolated atoms and diatomic molecules the coupling of two electronic states by an intense x-ray pulse of resonant frequency and the resulting Rabi oscillations of the population between these states have been recently studied  \cite{Rohringer08,Liu10,Sun10,Demekhin11SFatom,MolRaSfPRL,DemekhinCOarXiv}.

The process proposed here can take place in any atom and molecule which possesses accessible states with sufficient excess energy above the ionization potential of its environment to allow for ICD to occur. However, we would like to suggest a candidate suitable for experimental verification and at the same time this candidate should be amenable to high quality calculations including the nuclear dynamics of the whole process. The Ne dimer fulfills these prerequisites. We report computed data and ICD electron spectra which may directly be compared with experiments. Such an experiment to verify the proposed mechanism is feasible at present, i.e. at the FLASH facility.

\begin{figure}
\includegraphics[scale=0.40]{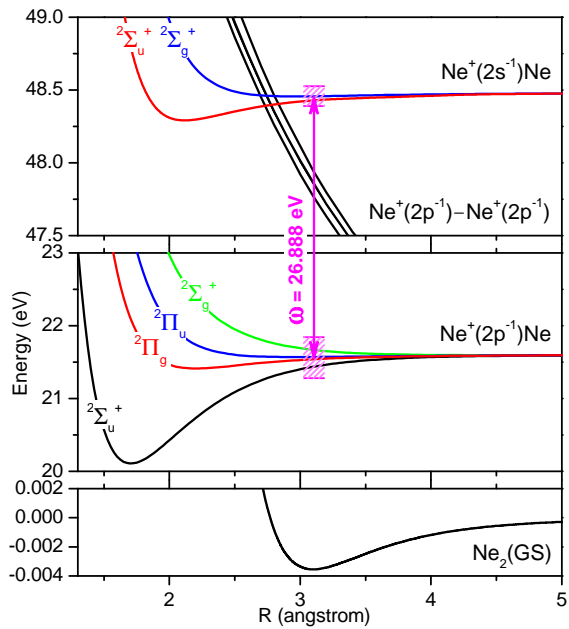}
\caption{(Color online) \emph{Ab initio} PECs of the states relevant to the process (\ref{eq:scheme}) of exposing the Ne dimer to an intense laser pulse. \emph{ Lower panel:} Of the Ne$_2$ ground state. \emph{Middle panel:}  PECs of the Ne$^+(2p^{-1})$Ne  ionic OV states. \emph{Upper panel:} PECs of the Ne$^+(2s^{-1})$Ne ionic  IV states which decay by ICD and of the dications Ne$^+(2p^{-1})$--Ne$^+(2p^{-1})$ which are the final states of ICD. The photon frequency of  $\omega= 26.888$~eV used in the calculations is indicated in the figure. Note the different energy scales in all panels.}\label{fig_pecs}
\end{figure}

For the example chosen the process can be schematically described as follows:
\begin{equation}
\label{eq:scheme}
\begin{array}{rl}
\mbox{Ne}_2~\stackrel{\omega}{\longrightarrow}    & \mbox{Ne}^+(2p^{-1})\mbox{Ne} + e_{ph} \\
& ~~~~~~~~\Updownarrow\text{\footnotesize{$\omega$}} \\
& \mbox{Ne}^+(2s^{-1})\mbox{Ne}  \\
&~~~~~~~~\downarrow\text{\tiny{\emph{ICD}}}\\
\multicolumn{2}{l}{ ~~~~~~~\mbox{Ne}^+(2p^{-1})+ \mbox{Ne}^+(2p^{-1}) +e_{ICD}} 
\end{array}
\end{equation}
The field ionizes a $2p$ electron of a Ne atom in the dimer and couples resonantly the resulting OV ionic states with the respective  IV ionic states.  The vertical double-arrow in (\ref{eq:scheme}) indicates the strong coupling between the OV and IV ionic states. Fig.~\ref{fig_pecs} depicts the \emph{ab initio} computed potential energy curves (PECs) of  the states relevant for the process. At the equilibrium internuclear distance ($R_e=3.1$~\AA) of the Ne dimer, the OV ionic states  Ne$^+(2p^{-1})$Ne (middle panel) are nearly degenerate and this holds also for the IV ionic states Ne$^+(2s^{-1})$Ne  (uppermost panel). The near degeneracy becomes rather perfect at larger internuclear distances. To resonantly couple the OV and IV states by an intense laser field we choose a pulse with a central frequency of  $\omega=26.888$~eV, see Fig.~\ref{fig_pecs}, which is the energy difference between the statistically weighted average of the two groups of computed energies at 3.1~\AA.  The IV ionic states shown in the upper panel of Fig.~\ref{fig_pecs} decay by ICD producing the strongly repelling Ne$^+(2p^{-1})$--Ne$^+(2p^{-1})$  states (also shown in this panel) which undergo a Coulomb explosion and low-energy ICD electrons. Importantly for experiment, the photoelectrons produced in the first step of (\ref{eq:scheme}) initiating the whole process have an energy of around 5.32~eV which is well separated from the energy of the ICD electrons.  Finally we stress that  Ne$^+(2p^{-1})$--Ne$^+(2p^{-1})$  can be competitively produced by the direct ionization of the neutral Ne atom in Ne$^+(2p^{-1})$Ne  produced in the first step of (\ref{eq:scheme}) and similarly Ne$^+(2s^{-1})$--Ne$^+(2p^{-1})$  (not shown in Fig.~\ref{fig_pecs}) by the direct ionization of  Ne$^+(2s^{-1})$Ne  resulting in the second step of (\ref{eq:scheme}). All of these states and processes are included in the computations.

\begin{figure}
\includegraphics[scale=0.30]{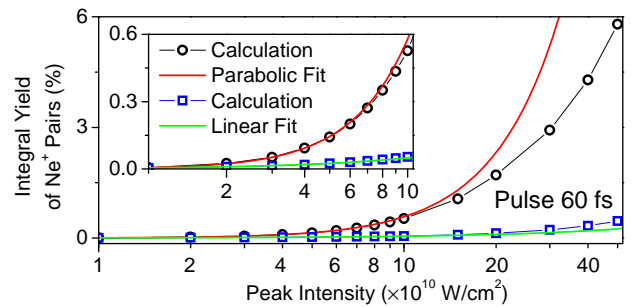}
\caption{(Color online)  Integral yield of the  Ne$^+$ ion pairs produced by exposing Ne dimers to a Gaussian laser pulse of 60 fs duration as a function of the peak intensity of the pulse. Open circles: Yield computed for the process (\ref{eq:scheme}) for a pulse with a central frequency $\omega= 26.888$~eV. At least two photons are required for the process. Open squares: Yield computed for a pulse with a central frequency of of $\omega=53.776$~eV. The ICD is triggered directly by the absorption of one photon. At the intensities shown the ICD initiated by process (\ref{eq:scheme}) via two-photon absorption is clearly the more efficient mechanism for the production of these ionic fragments.}\label{fig_intens}
\end{figure}

In order to compute the process~(\ref{eq:scheme}) we combine the previously developed theoretical and computational approaches \cite{Demekhin11SFatom,MolRaSfPRL,DemekhinCOarXiv} to evaluate the resonant Auger decay effect in intense laser fields (see \cite{SM} for details). The nuclear dynamics have been calculated employing the efficient Multi-Configuration Time-Dependent Hartree (MCTDH) method and code  \cite{Meyer90mctdh,MCTDH}. Thereby we have utilized the \emph{ab initio} PECs and the \emph{ab initio} ICD transition rates reported in Ref.~\cite{Averbukh06}. The values of the electron transition matrix elements were extracted from the experimental photoionization cross section of the Ne atom ($\sigma_{2p}={7.8}$~Mb at  28.4~eV \cite{BeckShirl}) and the experimental $2s^{-1}\to 2p^{-1}$  radiative decay rate of Ne$^+$  ($\Gamma_r=4.8~\mu$eV   \cite{Lablanquie00}  corresponding to the radiative lifetime of $\tau_r\sim 0.14$~ns).

\begin{figure}
\includegraphics[scale=0.35]{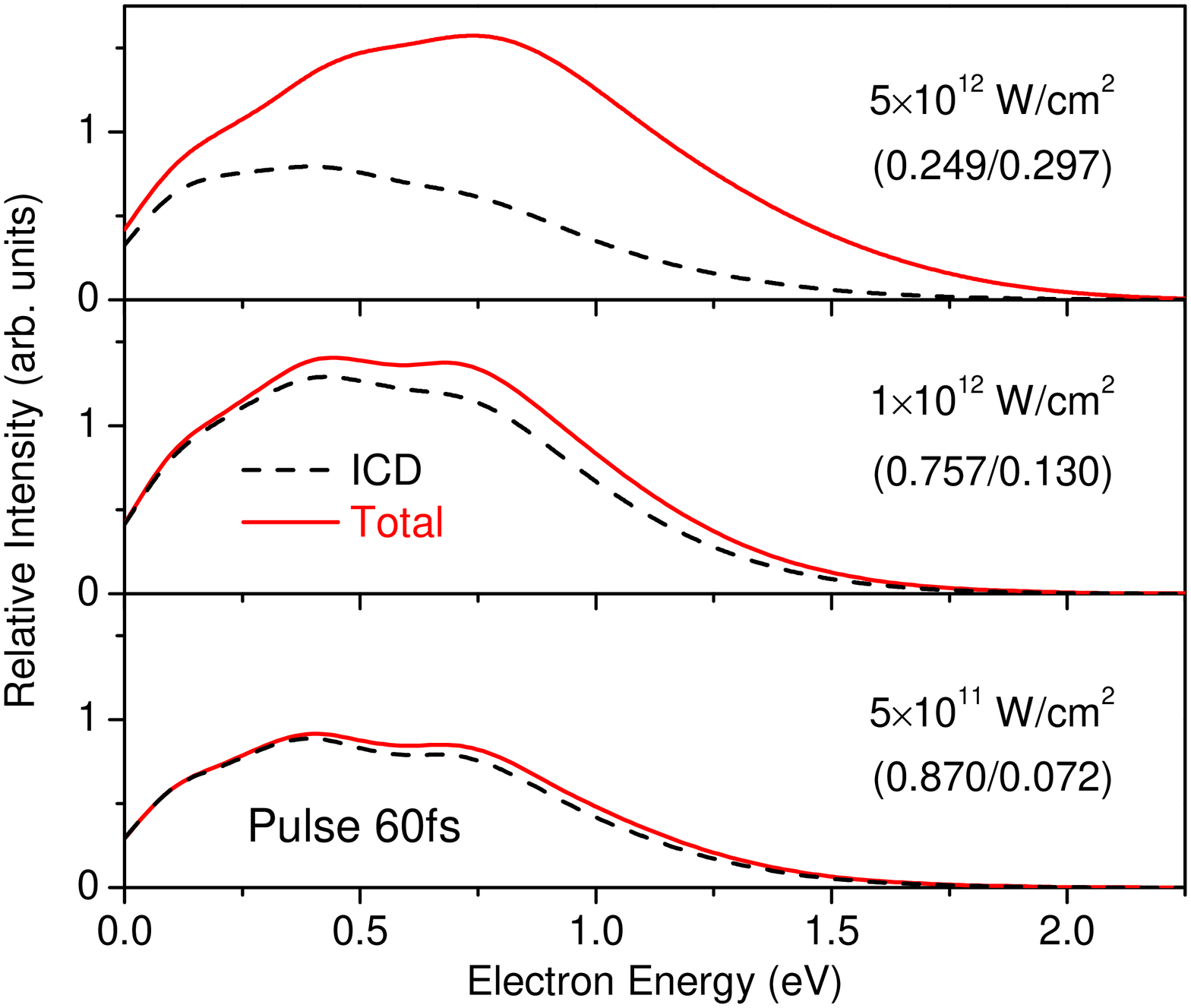}
\caption{(Color online) Electron spectra after exposure of Ne$_2$ to a 60~fs Gaussian pulse for three intensities (solid curves,  `Total').  For comparison the individual contributions of the ICD process (broken curves,  `ICD') are also shown.  For each intensity two numbers are reported in brackets: the fraction of neutral dimers which have survived the pulse, and the fraction of dimers which have become singly ionized.}\label{fig_60fs}
\end{figure}

Fig.~\ref{fig_intens} depicts the integral yield of the Ne$^+$ ion pairs produced by exposing the Ne dimer to a Gaussian-shaped pulse of 60~fs duration as a function of the pulse peak intensity. For comparison, we remind that the ICD lifetime is $\sim$80 fs \cite{Averbukh06}. The yield computed for the presently studied process (\ref{eq:scheme}) is depicted by open circles. At the relatively low intensities considered in the figure, the ICD initiated by two-photon absorption is by far the dominant mechanism for the production of these ionic fragments. As expected, the yield grows quadratically with the field intensity as the fit in Fig.~\ref{fig_intens} to the five points lowest in intensity shows. How does this yield compare with the yield obtained in the traditional way of initiating ICD by the absorption of a single-photon of energy above the IV ionization threshold? To answer this basic question we also computed the ICD triggered directly by a strong pulse with twice the central frequency. The results are shown as open squares in the figure. The computed yield grows linearly with the intensity, as the fit in Fig.~\ref{fig_intens} to the first five points lowest in intensity indicates. Interestingly, the process (\ref{eq:scheme}) is much more efficient at the considered intensities, although it requires the absorption of at least two photons. This finding gives further motivation to investigate interatomic and intermolecular processes by intense lasers. 

At intensities larger than $10^{11}~\mathrm{W/cm}^2$ the computed integral yield starts to deviate from the fit. Indeed, as the field intensity grows other mechanisms than ICD induced by the field start to play an important role. As demonstrated in Refs.~\cite{Liu10,Sun10,Demekhin11SFatom}, the (direct) ionization of all the states participating in the ICD process results in losses (leakages) of population of these states. Particularly relevant and interesting is the direct ionization of Ne$^+(2p^{-1})$Ne  mentioned above. This ionization and the ICD populate the  Ne$^+(2p^{-1})$--Ne$^+(2p^{-1})$ states coherently, thus naturally inducing interference effects in the resulting electron spectra \cite{Demekhin11SFatom}. As has been discussed and demonstrated in Refs.~\cite{MolRaSfPRL,DemekhinCOarXiv}, an intense field also gives rise to light-induced non-adiabatic effects. All these mechanisms were simultaneously incorporated in the present calculations.

\begin{figure}
\includegraphics[scale=0.30]{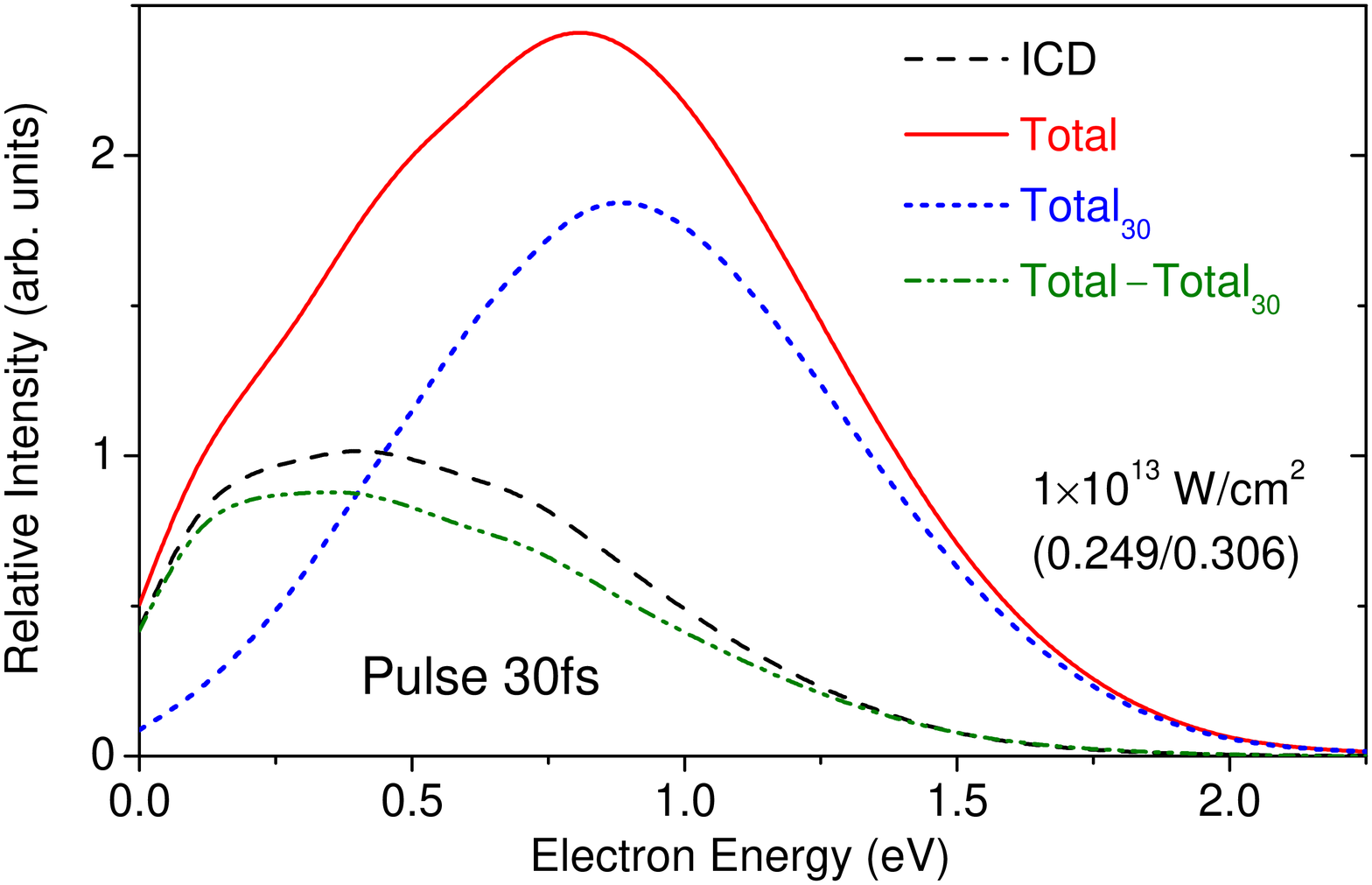}
\caption{(Color online) Electron spectrum after exposure of Ne$_2$ to a 30 fs Gaussian pulse of intensity $1\times10^{13}$~W/cm$^2$ (solid curve,  `Total'). Also shown are the individual contribution of the ICD process (dashed curve,  `ICD'), the spectrum measured 30~fs after the pulse maximum (dotted curve,  `Total$_{30}$'), and the difference between the `Total' and the `Total$_{30}$' spectra (dash-dotted curve). }\label{fig_30time}
\end{figure}

Let us now turn to the spectra of the electrons emitted in the process (\ref{eq:scheme}). The electron spectra computed for Gaussian-shaped 60 fs pulses of three different peak intensities are depicted in Fig.~\ref{fig_60fs}  by solid curves (labeled `Total'). Although the intensities chosen are not the strongest FEL pulse intensities available at present, they cannot be accessed by any conventional synchrotron sources. After the pulse is over and all the IV states populated by it have decayed via ICD, a fraction of the neutral dimers, of course, survives and remains in their ground state. For each intensity this fraction is reported in the figure (first number in the bracket below the indicated intensity). Also of interest are the corresponding fractions of the dimers which have become singly ionized Ne$_2^+$ dimers. These numbers are reported  in the figure next to the neutral fraction. At the lowest considered peak intensity ($5\times10^{11}$~W/cm$^2$), the pulse manages to ionize 13\% of the neutral dimers. Of these, more than half  (7.2\%) remain as Ne$_2^+$  ions and the rest  (5.8\%) participate further in the ICD. Even at highest considered intensity of $5\times10^{12}$~W/cm$^2$, the ionization of the ground and of the ionic states are still far from being saturated: about a quarter of the neutral dimers survive, and of the three quarters which have been ionized about 30\% remain as Ne$_2^+$ ions and  45\% have undergone further ionization and have emitted an electron which contributes to the spectrum shown.  The calculations indicate that sequential single-photon ionization dominates over simultaneous two-photon absorption.

As discussed above, not all of the low-energy electrons in the computed spectra in Fig.~\ref{fig_60fs} are produced via the ICD. In order to reveal the individual contribution of the ICD process, we performed additional  calculations. In these all the pathways to produce Ne$^+$--Ne$^+$ dicationic states by direct ionization of Ne$_2^+$ ions were excluded, but all losses from the neutral and ionic states were still taken into account. The results of these approximate calculations are depicted in Fig.~\ref{fig_60fs} by broken curves (labeled `ICD'). The difference between the solid and broken curves at each peak intensity represents the contribution to the spectrum deriving from the direct ionization pathways and the interference between the direct and ICD pathways.

\begin{figure}
\includegraphics[scale=0.30]{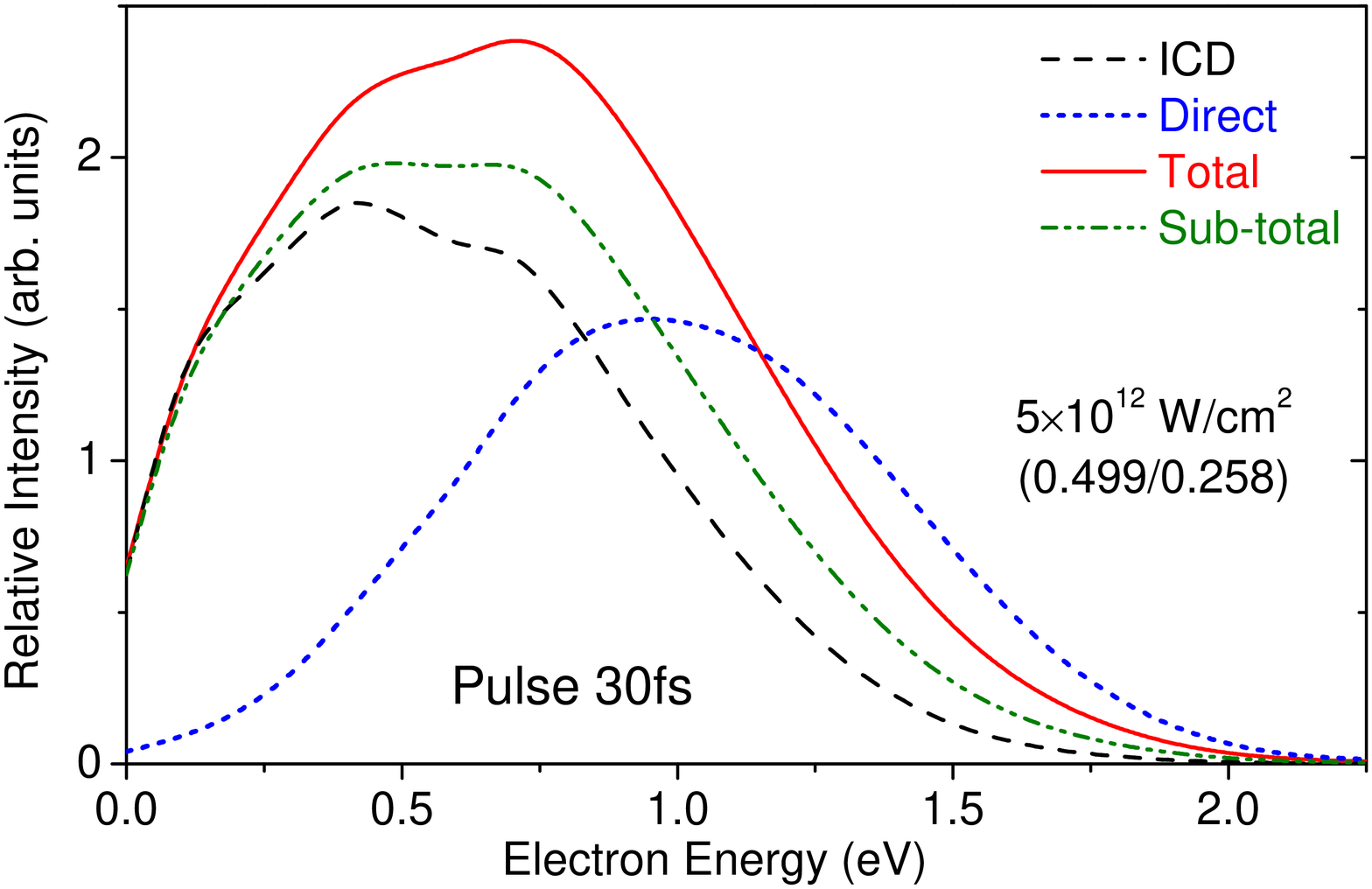}
\caption{(Color online) Electron spectrum after exposure of Ne$_2$ to a 30 fs Gaussian pulse of intensity $5\times10^{12}$~W/cm$^2$ (solid curve,  `Total'). For analysis the individual contribution of the ICD process (dashed curve,  `ICD'), the individual contribution of the direct ionization, i.e., excluding ICD from the calculations (dotted curve,  `Direct'), and the results obtained excluding only the ionization of Ne$^+(2s^{-1})$Ne to produce Ne$^+(2s^{-1})$--Ne$^+(2p^{-1})$ (dashed-dotted curve,  `Sub-total') are shown.}\label{fig_30appr}
\end{figure}

As seen in Fig.~\ref{fig_60fs}, almost all low-energy electrons are produced via the ICD for intensities below $5\times10^{11}$~W/cm$^2$. Interestingly, the shape of the ICD spectrum computed for this peak intensity is rather similar to that of the experimental ICD spectrum measured in Ref.~\cite{Jahnke04} by synchrotron radiation in the single-photon absorption mode. ICD is still by far the dominant pathway for production of low-energy electrons at $1\times10^{12}$~W/cm$^2$ (middle panel), but the direct ionization pathways become more and more relevant as the intensity increases. This leads to a change of the shape of the electron spectrum, indicating already now (see below) that the direct pathways preferentially produce electrons at somewhat higher energies than ICD.

How to identify at strong field the individual contribution of ICD to the spectrum?  We may exploit the fact that the two kinds of pathways for the production of the final ionic fragments take place on different timescales. Obviously, the direct ionizations are operative only when the pulse is on. The ICD process itself is, in turn, independent of the pulse duration and in the present example the ICD lifetime is $\sim 80$~fs \cite{Averbukh06}. In order to separate the two kinds of pathways, one has to apply shorter pulses and invoke measurements in the time domain.

To demonstrate this idea we show in Fig.~\ref{fig_30time} electron spectra computed for a Gaussian pulse of half the duration (30~fs) and twice the intensity ($1\times10^{13}$~W/cm$^2$)  of  the pulse used to obtain the spectrum in the upper panel of Fig.~\ref{fig_60fs}. As usual, the total spectrum (solid curve) is obtained at long time after the process is over. Measuring the spectrum after a time delay of  30~fs from the pulse maximum, i.e., just after the pulse expired, produces the spectrum labeled  `Total$_{30}$' depicted as a dotted curve. During the time the pulse is on the direct ionization is expected to strongly dominate the production of the spectrum and the difference between the `Total' and the `Total$_{30}$' spectra is thus expected to reflect the individual contribution of the ICD. This difference, shown in Fig.~\ref{fig_30time} as a dash-dotted  curve, is indeed very similar to the spectrum computed excluding the direct ionization pathways (dashed curve). This finding confirms that most ICD electrons are emitted after the pulse has expired.

It is evident from Figs.~\ref{fig_60fs} and \ref{fig_30time} that direct photoionization and ICD preferentially produce electrons in different energy ranges. This finding is further analyzed in Fig.~\ref{fig_30appr} for a Gaussian pulse of 30~fs duration and $5\times10^{12}$~W/cm$^2$ peak intensity. The individual contribution of the direct ionization is shown by the dotted curve (labeled `Direct'), and that of  ICD by the dashed curve. Clearly, ICD mainly contributes to the electron energy range 0--1 eV, and the direct channel produces electrons mainly between 0.5 and 1.5~eV. Most importantly, the two contributions are by no means additive and the strong interference between the pathways determines the final shape of the spectrum.  As a last point we mention that at high intensities also the ionization of  Ne$^+(2s^{-1})$Ne  which can decay by ICD contributes and cannot be neglected. At least a third photon must be absorbed in the process for this ionization which produces Ne$^+(2s^{-1})$--Ne$^+(2p^{-1})$. The impact of this ionization can be seen in Fig.~\ref{fig_30appr} (dash-dotted curve labelled `Sub-total').

In conclusion, a general mechanism to initiate ICD by an intense laser pulse is proposed. The initiation requires the absorption of at least two-photons. Nevertheless, it is found here to initiate interatomic processes much more efficiently than by utilizing traditional one-photon ionization schemes with photons of sufficiently high energies. The process is exemplified by extensive calculations on the Ne dimer which take account of losses via ionization induced by the intense field and incorporate the quantum nuclear dynamics on all the participating electronic levels including their coupling by the pulse. At moderate field intensities, ICD is the main mechanism for the production of low-energy electrons. At higher intensities other ionization mechanisms become relevant as well. Importantly, they coherently interfere with ICD enriching the process. Since these other mechanisms are only operative while the pulse is on, there is an interesting interplay between them and ICD which can be controlled by varying the pulse. In particular, pump-probe techniques can be used to identify the contribution of ICD to the electron spectrum even at higher intensity by a single additional measurement after a time delay. It is our opinion that the conclusions drawn are rather general and apply to many systems. We hope that the present results will stimulate experiments.

\begin{acknowledgements}
The research leading to these results has received funding from the ERC under the EU's  FP7,  AIG No. 227597. 
\end{acknowledgements}


\begin{thebibliography}{38}
\expandafter\ifx\csname natexlab\endcsname\relax\def\natexlab#1{#1}\fi
\expandafter\ifx\csname bibnamefont\endcsname\relax
  \def\bibnamefont#1{#1}\fi
\expandafter\ifx\csname bibfnamefont\endcsname\relax
  \def\bibfnamefont#1{#1}\fi
\expandafter\ifx\csname citenamefont\endcsname\relax
  \def\citenamefont#1{#1}\fi
\expandafter\ifx\csname url\endcsname\relax
  \def\url#1{\texttt{#1}}\fi
\expandafter\ifx\csname urlprefix\endcsname\relax\def\urlprefix{URL }\fi
\providecommand{\bibinfo}[2]{#2}
\providecommand{\eprint}[2][]{\url{#2}}

\bibitem{Cederbaum97prlicd}
L.S. Cederbaum, \emph{et al.}, Phys. Rev. Lett. \textbf{79},   4778 (1997).

\bibitem{Averbukh}
V. Averbukh, \emph{et al.}, J. Electr. Spectr. Relat. Phen.  \textbf{183}, 36 (2011).

\bibitem{Hergenhahn}
U. Hergenhahn, J. Electr. Spectr. Relat. Phen.  \textbf{184}, 78 (2011).

\bibitem{Marburger03}
S. Marburger,  \emph{et al.}, Phys. Rev. Lett. \textbf{90}, 203401 (2003).

\bibitem{Jahnke04}
T. Jahnke, \emph{et al.}, Phys. Rev. Lett. \textbf{93}, 163401 (2004).

\bibitem{Morishita06}
Y. Morishita, \emph{et al.}, Phys. Rev. Lett. \textbf{96}, 243402 (2006). 

\bibitem{Jahnke10}
T. Jahnke, \emph{et al.}, Nature Physics \textbf{6}, 139 (2010).

\bibitem{Mucke10}
M. Mucke, \emph{et al.}, Nature Physics \textbf{6}, 143 (2010).

\bibitem{Aziz08}
E.F. Aziz, \emph{et al.}, Nature \textbf{455}, 89 (2008).

\bibitem{Kryzhevoi11}
N.V. Kryzhevoi and L.S. Cederbaum, Angew. Chem. Int. Ed. \textbf{50}, 1306 (2011).

\bibitem{Stoychev11}
S.D.\,Stoychev,\,\emph{et\,al.},\,J.\,Am.\,Chem.\,Soc.\,\textbf{133},\,6817\,(2011).  

\bibitem{FLASH}
W. Ackermann, \emph{et al.},  Nature photonics \textbf{1}, 336  (2007).

\bibitem{XFEL1}
P. Emma, \emph{et al.},  Nature photonics \textbf{4}, 641  (2010).

\bibitem{XFEL2}
L. Young, \emph{et al.},  Nature \textbf{466}, 56 (2010).

\bibitem{Saalmann06}
U. Saalmann, \emph{et al.}, J. Phys. B \textbf{39}, R39 (2006).

\bibitem{Averbukh09}
V. Averbukh and P. Koloren\v{c}, Phys. Rev. Lett. \textbf{103}, 183001 (2009).

\bibitem{Kuleff10}
A.I. Kuleff, \emph{et al.}, Phys. Rev. Lett. \textbf{105}, 043004 (2010).

\bibitem{Rohringer08}
N.\,Rohringer\,and\,R.\,Santra,\,Phys.\,Rev.\,A\,\textbf{77},\,053404\,(2008).

\bibitem{Liu10}
 J.-C. Liu,  \emph{et al.}, Phys. Rev. A \textbf{81},  043412  (2010).

\bibitem{Sun10}
Y.-P. Sun,  \emph{et al.}, Phys. Rev. A \textbf{81},  013812  (2010).

\bibitem{Demekhin11SFatom}
Ph.V. Demekhin and L.S. Cederbaum, Phys. Rev. A  \textbf{83},  023422  (2011).

\bibitem{MolRaSfPRL}
L.S.\,Cederbaum,\,\emph{et\,al.},\,Phys.\,Rev.\,Lett.\,\textbf{106},\,123001\,(2011).

\bibitem{DemekhinCOarXiv}
Ph.V. Demekhin, \emph{et al.}, arXiv:1105.5374.

\bibitem{SM}
Supplemental material for this manuscript

\bibitem{Meyer90mctdh}
H.-D.Meyer, \emph{et al.},  Chem. Phys. Lett. \textbf{165}, 73  (1990).

\bibitem{MCTDH}
G.A. Worth, M.H. Beck,  A. J\"{a}ckle, and H.-D. Meyer. The MCTDH Package, see http://mctdh.uni-hd.de.

\bibitem{Averbukh06}
V. Averbukh  and L.S. Cederbaum, J. Chem. Phys. \textbf{125}, 094107 (2006)

\bibitem{BeckShirl}
U.~Becker and D.~Shirley, \emph{{VUV} and {S}oft {X}-{R}ay {P}hotoionization}, 
(\bibinfo{publisher}{Plenum Press}, \bibinfo{address}{New York},
  \bibinfo{year}{1996}),  pp. 135--180.

\bibitem{Lablanquie00}
P. Lablanquie, \emph{et al.}, Phys. Rev. Lett. \textbf{84}, 431 (2000).

\end{thebibliography}
\end{document}


\title{Supplemental Material\\ ~\\
Exploring  Interatomic Coulombic Decay by Free Electron Lasers}

\author{\firstname{Philipp~V.} \surname{Demekhin}}
\affiliation{Theoretische Chemie, Physikalisch-Chemisches
Institut, Universit\"{a}t  Heidelberg, Im Neuenheimer Feld 229,
D-69120 Heidelberg, Germany}\email{philipp.demekhin@pci.uni-heidelberg.de}

\author{\firstname{Spas D.} \surname{Stoychev}}
\affiliation{Theoretische Chemie, Physikalisch-Chemisches Institut, Universit\"{a}t Heidelberg,
Im Neuenheimer Feld 229, D-69120 Heidelberg, Germany}

\author{\firstname{Alexander I. } \surname{Kuleff}}
\affiliation{Theoretische Chemie, Physikalisch-Chemisches Institut, Universit\"{a}t Heidelberg,
Im Neuenheimer Feld 229, D-69120 Heidelberg, Germany}

\author{\firstname{Lorenz~S.} \surname{Cederbaum}}
\affiliation{Theoretische Chemie, Physikalisch-Chemisches Institut, Universit\"{a}t Heidelberg,
Im Neuenheimer Feld 229, D-69120 Heidelberg, Germany}

\pacs{33.20.Xx, 32.80.Hd, 41.60.Cr, 82.50.Kx}

\maketitle

\section{Theory: nuclear dynamics hamiltonian}

In this supplementary information we provide an extended representation of the presently applied theoretical approach. It implies a general formulation of the time-dependent theory for the nuclear wave packet propagation (see, e.g., Refs.~[S1--S5]), which has recently been extended and applied to evaluate the  resonant Auger decay effect of atoms and molecules in intense laser fields [S6--S8]. The extended  approach and all necessary derivations of the theory can be found in Refs.~[S5--S8]. In order to make it easier for the interested reader to follow our approach without restoring to the literature, its  essential points and `working' equations describing the presently studied process  are collected and discussed below.

We start with the schematic representation of the example chosen in the main text:
\begin{equation}
\label{eq:scheme}
\begin{array}{rl}
\mbox{Ne}_2~\stackrel{\omega}{\longrightarrow}    & \mbox{Ne}^+(2p^{-1})\mbox{Ne} + e_{ph} \\
& ~~~~~~~~\Updownarrow\text{\footnotesize{$\omega$}} \\
& \mbox{Ne}^+(2s^{-1})\mbox{Ne}  \\
&~~~~~~~~\downarrow\text{\tiny{\emph{ICD}}}\\
\multicolumn{2}{l}{ ~~~~~~~\mbox{Ne}^+(2p^{-1})+ \mbox{Ne}^+(2p^{-1}) +e_{ICD}} 
\end{array}
\end{equation}
The strong laser field ionizes a 2p electron of a Ne atom in the dimer populating outer-valence (OV) ionized states  Ne$^+(2p^{-1})$Ne and producing a  photoelectron $e_{ph}$. The same pulse with the resonant central frequency $\omega=26.888$~eV couples also the resulting OV and inner-valence (IV) ionized states Ne$^+(2s^{-1})$Ne, as indicated by the vertical double arrow in Eq.~(\ref{eq:scheme}). The IV ionic states of the  dimer   relax by interatomic Coulombic decay (ICD) producing the strongly repelling $\mbox{Ne}^+(2p^{-1})-\mbox{Ne}^+(2p^{-1})$  states which undergo a Coulomb explosion and low-energy ICD electrons $e_{ICD}$. The $\mbox{Ne}^+(2p^{-1})-\mbox{Ne} ^+(2p^{-1})$ states can be competitively produced by the direct ionization of the neutral Ne atom in Ne$^+(2p^{-1})$Ne states produced in the first step of (\ref{eq:scheme}). Similarly,  the  ionization of IV ionized states  Ne$^+(2s^{-1})$Ne resulting in the second step of (\ref{eq:scheme}) to directly produce the $\mbox{Ne}^+(2s^{-1})-\mbox{Ne}^+(2p^{-1})$ states cannot be neglected at high intensities of the field.  All of these states and processes are included in the present computations.

In order to theoretically describe the dynamics of the process (\ref{eq:scheme})  as a function of time, we solve the time-dependent Schr\"{o}dinger equation for the Ne$_2$ and its interaction with the field (atomic units $e=m_e=\hbar=1$ are used throughout)
\begin{equation}
\label{eq:hamilt}
i\dot{\Psi}(t)=\hat{H}(t)\Psi(t)=\left(\hat{H}_{nuc}+\hat{H}_{el}+ \hat{D}\, \mathcal{E}(t)\right)\Psi(t) . 
\end{equation} 
where $\hat{D}$ represents the dipole transition operator.  We assume a  coherent and monochromatic Gaussian-shaped pulse of   duration $\tau$ centered at $t_0$ ($g(t)=e^{-(t-t_0)^2/\tau^2}$) with linear polarization of the field along the $z$ axis and  central frequency $\omega$:
\begin{equation}
\label{eq:e_vector}
\mathcal{E}(t)=\mathcal{E}_0(t)\cos\omega t=\mathcal{E}_0 \,g(t) \cos\omega t.
\end{equation}
Here $\mathcal{E}_0$ is the peak amplitude, and the pulse-shape function $g(t)$   varies slowly on the timescale of $2\pi/\omega$. The cycle-averaged intensity of the field is given in atomic units via (1 a.u. = 6.43641$\times 10^{15}$ W/cm$^2$)
\begin{equation}
\label{eq:intens}
I(t)=\frac{1}{8\pi\alpha}\left\{\mathcal{E}_0 \,g(t)\right\}^2 ,
\end{equation}
where $ \alpha=1/137.036$ is the fine structure constant.

Following Refs.~[S5--S8] the total wave function of a rotating molecule as a function of time can be represented via the  ansatz including all  electronic states participating in the process~(\ref{eq:scheme}). It includes: the ground state of the Ne$_2$ with the electronic wave function  $\Phi_I$; all intermediate  Ne$^+(2p^{-1})$Ne (OV-ionized) and  Ne$^+(2s^{-1})$Ne  (IV-ionized)  states  plus outgoing  photoelectron of energy $\varepsilon_{ph}$ with the total electronic wave functions $\Phi_{OV}^{\varepsilon_{ph}}$ and $\Phi_{IV}^{\varepsilon_{ph}}$;    and all final  $\mbox{Ne}^+(2p^{-1})-\mbox{Ne}^+(2p^{-1})$ (OVOV doubly-ionized) states plus outgoing photoelectron and ICD electron of  energy $\varepsilon_{ICD}$ with the total electronic wave function $\Phi_{OVOV}^{\varepsilon_{ph}\varepsilon_{ICD}}$
\begin{multline}
\label{eq:anzatz}
\Psi(t)=  \Psi_I(t)  \Phi_I +\sum_j\int   \widetilde{\Psi}_{OV_j}  (\varepsilon_{ph},t)    \Phi_{OV_j}^{\varepsilon_{ph}}  d\varepsilon_{ph} +\sum_j\int   \widetilde{\Psi}_{IV_j}  (\varepsilon_{ph},t)    \Phi_{IV_j}^{\varepsilon_{ph}}  d\varepsilon_{ph} +\\
+\sum_j\iint    \widetilde{\Psi}_{OVOV_j} (\varepsilon_{ph},\varepsilon_{ICD},t)    \Phi_{OVOV_j}^{\varepsilon_{ph}\varepsilon_{ICD}}  d\varepsilon_{ph} d\varepsilon_{ICD},
\end{multline}
where index $j$ numerates all possible states within each subset of states. The functions  $ \Psi_I(t) $, $  \widetilde{\Psi}_{OV_j}  (\varepsilon_{ph},t)  $, $   \widetilde{\Psi}_{IV_j}  (\varepsilon_{ph},t)   $ and $  \widetilde{\Psi}_{OVOV_j} (\varepsilon_{ph},\varepsilon_{ICD},t)  $ are the time-dependent wave packets propagating on the potential energy surfaces  of the included electronic states. It should be remembered that these wave packets depend  on the nuclear  coordinates $R$ and $\theta$. For brevity, we will explicitly  show the  dependence on the nuclear coordinates only in the final Hamiltonian matrix (\ref{eq:fin2}) governing the nuclear dynamics of the process~(\ref{eq:scheme}).

In order to obtain the final set of equations for the propagation of the nuclear wave packets, we substitute the ansatz  (\ref{eq:anzatz}) in the time-dependent  Schr\"{o}dinger equation (\ref{eq:hamilt}) and project the result onto each electronic state. As explicitly demonstrated in Refs.~[S5--S8], in the derivation  we imply the rotating wave approximation [S9], the local approximation [S10,S11], and redefine (`dress') the time dependent wave packets  as follows
\begin{subequations}
\label{eq:redefine}
\begin{equation}
   {\Psi}_{OV_j}  (\varepsilon_{ph},t)  =   \widetilde{\Psi}_{OV_j}  (\varepsilon_{ph},t)  \,e^{+i\omega t}
\end{equation}
\begin{equation}
   {\Psi}_{IV_j}  (\varepsilon_{ph},t)  =   \widetilde{\Psi}_{IV_j}  (\varepsilon_{ph},t)  \,e^{+2i\omega t}.
\end{equation}
\begin{equation}
   {\Psi}_{OVOV_j} (\varepsilon_{ph},\varepsilon_{ICD},t)  =    \widetilde{\Psi}_{OVOV_j} (\varepsilon_{ph},\varepsilon_{ICD},t)  \,e^{+2i\omega t}
\end{equation}
\end{subequations}
Of course, all electronic states within the subsets of IV and OV singly-ionized, and OVOV doubly-ionized final states shown in Fig.~1 of the main text participate in the dynamics of the considered process and have to be included in the  ansatz (\ref{eq:anzatz}). For the sake of transparency of   presentation  we explicitly write  in  the ansatz (\ref{eq:anzatz}) below   only one electronic state from each of the subsets. In particular, we show  only the Ne$^+(2p^{-1})$Ne  OV-ionized state of $^2\Pi_g$ symmetry  coupled by the field with the Ne$^+(2s^{-1})$Ne  IV-ionized state of $^2\Sigma^+_u$ symmetry, which subsequently decays via ICD into one of the OVOV doubly-ionized final states. The extension of the expressions  to the case of all participating states is straightforward.   

Let us collect the individual  nuclear wave packets contributing to the total wave function (\ref{eq:anzatz}) into a single vector
\begin{equation}
\label{eq:fin1}
\vert \overline{\Psi}(\varepsilon_{ph},\varepsilon_{ICD},t)\rangle= \left(\begin{array}{l} \vert\Psi_I(t)\rangle  \\ \vert {\Psi}_{OV}  (\varepsilon_{ph},t) \rangle\\  \vert {\Psi}_{IV}  (\varepsilon_{ph},t) \rangle \\  \vert {\Psi}_{OVOV} (\varepsilon_{ph},\varepsilon_{ICD},t) \rangle \end{array} \right),
\end{equation}
and define the matrix Hamiltonian
\begin{multline}
\label{eq:fin2}
\hat{\mathbf{H}}(R,\theta,t)= \hat{\textbf{T}}(R,\theta)+ \\ 
\left(\begin{array}{l|l|l|l} V_I(R)-\frac{i}{2}\Gamma_I^{ph}(t)~ &0&0&0\\ \hline
d_x(t)\sin \theta +  &V_{OV}(R)-\frac{i}{2}\Gamma_{OV}^{ph}(t)+  & \left(D^\dag_x(t)-\frac{i}{2} W^\dag(t)\right) \sin \theta ~&0\\
+d_z(t)\cos \theta &+\varepsilon_{ph}-\omega\ & &\\ \hline
0&\left(D_x(t)-\frac{i}{2} W(t)\right) \sin \theta~  & V_{IV}(R) +\varepsilon_{ph}-2\omega -& 0\\
& &  -\frac{i}{2}[\Gamma^{ICD}_{IV}(R)+\Gamma_{IV}^{ph}(t)] & \\ \hline
0& \widetilde{d}_x(t)\sin \theta +  & V_{ICD}(R)& V_{OVOV}(R)+ \\
&  +\widetilde{d}_z(t)\cos \theta  & &+\varepsilon_{ph}+\varepsilon_{ICD}  -2\omega \\
 \end{array} \right).
\end{multline}
With these notations, the final set of equations describing the propagation of the nuclear wave packets (\ref{eq:fin1})  now takes on  the following compact form
\begin{equation}
\label{eq:fin3}
i \vert \dot{\overline{\Psi}}(\varepsilon_{ph},\varepsilon_{ICD},t)\rangle = \hat{\mathbf{H}}(t) \,  \vert \overline{\Psi}(\varepsilon_{ph},\varepsilon_{ICD},t)\rangle.
\end{equation}
The matrix $\hat{\mathbf{H}} $ can be viewed as the effective Hamiltonian governing the two-dimensional nuclear  dynamics in the process~(\ref{eq:scheme})   in an intense laser field. Below we summarize the physical meaning of each term of the Hamiltonian matrix (\ref{eq:fin2}) and provide their explicit expressions.

$\hat{\textbf{T}}(R,\theta)$  is the common nuclear kinetic energy operator for the vibrational motion along the internuclear distance $R$ and the rotational motion described by the angle $\theta$ between the polarization vector of the laser pulse and the molecular axis. The functions  $V_I(R)$, $V_{OV}(R)$, $V_{IV}(R)$, and $V_{OVOV}(R)$ on the diagonal are the    potential energies of the corresponding electronic states. These potential curves  of the Ne$_2$ were computed  \emph{ab-initio} in the present work as described in Refs. [S12,S13]  and are depicted  in Fig.~1 of the main text.  The couplings between the various electronic states  are given by the following   matrix elements of the total Hamiltonian (\ref{eq:hamilt}) 
\begin{subequations}
\label{eq:couplings}
\begin{equation}
\label{eq:couplings_ion}
\langle \Phi_{OV}^{\varepsilon_{ph}}\vert \hat{H}(t) \vert \Phi_I\rangle= \left[d_x\sin \theta +   d_z\cos \theta  \right] \frac{\mathcal{E}_0 \,g(t)}{2}\, e^{-i\omega t}= \left[d_x(t)\sin \theta +   d_z(t)\cos \theta \right] \, e^{-i\omega t},
\end{equation}
\begin{equation}
\label{eq:couplings_exc}
\langle \Phi_{IV}^{\varepsilon_{ph}}\vert \hat{H}(t) \vert \Phi_{OV}^{\varepsilon_{ph}}\rangle=  D_x \sin \theta \, \frac{\mathcal{E}_0 \,g(t)}{2}\, e^{-i\omega t}= D_x(t)\sin \theta \, e^{-i\omega t},
\end{equation}
\begin{equation}
\label{eq:couplings_dir}
\langle \Phi_{OVOV}^{\varepsilon_{ph}\varepsilon_{ICD}}\vert \hat{H}(t) \vert \Phi_{OV}^{\varepsilon_{ph}} \rangle= \left[\widetilde{d}_x\sin \theta + \widetilde{d}_z\cos \theta  \right] \frac{\mathcal{E}_0 \,g(t)}{2}\, e^{-i\omega t}= \left[\widetilde{d}_x(t)\sin \theta +   \widetilde{d}_z(t)\cos \theta \right]  e^{-i\omega t},
\end{equation}
\begin{equation}
\label{eq:couplings_dec} 
\langle\Phi_{OVOV}^{\varepsilon_{ph}\varepsilon_{ICD}} \vert \hat{H}(t) \vert \Phi_{IV}^{\varepsilon_{ph}}\rangle= V_{ICD}(R),
\end{equation}
\end{subequations}
In Eqs.~(\ref{eq:couplings_ion}--\ref{eq:couplings_dir}), the   rotating wave approximation [S9]  has already been utilized, and, in contrast to the rapidly oscillating factor $e^{-i\omega t}$, the functions  $D(t)$ and $d(t)$ vary  slowly on the timescale of $2\pi/\omega$. Let us discuss these transition matrix elements.

The matrix element (\ref{eq:couplings_ion}) describes the ionization of a $2p$ electron of a Ne atom of the dimer in the ground electronic state via the $x$ and $z$ components of the dipole transition operator to produce OV-ionized state  Ne$^+(2p^{-1})$Ne and a photoelectron $\varepsilon_{ph}$. The matrix element (\ref{eq:couplings_exc}) represents the excitation of a 2s electron of Ne$^+$ ion into the vacant 2p orbital via the $x$ component of the dipole transition operator and couples, thereby, the selected  OV~$^2\Pi_g$ and IV~$^2\Sigma^+_u$  states of the singly-ionized dimer.  The matrix element (\ref{eq:couplings_dir}) is similar to  (\ref{eq:couplings_ion}) and corresponds to the ionization of the neutral Ne atom in the Ne$^+(2p^{-1})$Ne state to directly produce the $\mbox{Ne}^+(2p^{-1})-\mbox{Ne}^+(2p^{-1})$ final doubly-ionized   state and ICD electron $\varepsilon_{ICD}$. Eq.~(\ref{eq:couplings_dec}) defines the matrix element for the ICD transition, i.e. of the decay of the selected IV-ionized state into the OVOV  doubly-ionized final state   with   emission of the ICD electron via the Coulomb operator (note its explicit dependence on the internuclear distance $R$ [S14,S15]).

The ionization of the neutral dimers in the first step of the process~(\ref{eq:scheme}) transfers the  population from the ground electronic state  to the  OV-ionized states of the dimer by the transition  matrix element (\ref{eq:couplings_ion}) (matrix element $H_{21}$ in (\ref{eq:fin2})). The leakage of the corresponding population  from the ground state due to   direct photoionization into all possible final ionic states is described by the imaginary time-dependent  term $-\frac{i}{2}\Gamma_I^{ph}(t)$ on the respective  diagonal  ($H_{11}$) in the  Hamiltonian~(\ref{eq:fin2}). Its explicit expression was obtained in [S6] in the local approximation  and reads:
\begin{equation}
\label{eq:solut_gph}
\Gamma_I^{ph}(t)= 2\pi \sum _j  \vert d_{x/z}^j(t)\vert^2,
\end{equation}
where $d_{x/z}^j(t)$ are the dipole transition matrix elements (\ref{eq:couplings_ion}) for the direct ionization of the ground electronic state into all possible final ionic states numerated by superscript $j$. The total probability for the direct photoionization of the ground state (\ref{eq:solut_gph}) is identical to the quantity  $\gamma^{ph}(t)$ introduced in Ref.~[S16]:
\begin{equation}
\label{eq:solut_gelm}
\Gamma_I^{ph}(t)=\gamma^{ph}(t)=\sigma_{I}^{ph}I(t)/\omega,
\end{equation}
where $\sigma_{I}^{ph}$ is the total direct photoionization cross section of the ground state for the exciting-photon energy $\omega$, $I(t)$ is the field intensity (\ref{eq:intens}), and the quantity $I(t)/\omega$ stands in Eq.~(\ref{eq:solut_gelm}) for the photon flux.

Owing to the redefinition (\ref{eq:redefine}), the  potential energies of the IV and OV ionized states with a photoelectron on the respective diagonal elements of $\hat{\mathbf{H}}$   are `dressed'  by the field (i.e., $V_{OV}(R)+\varepsilon_{ph}-\omega$ and $V_{IV}(R) +\varepsilon_{ph}-2\omega$). They are also augmented by time-dependent imaginary terms  $-\frac{i}{2}\Gamma_{OV}^{ph}(t)$ and $-\frac{i}{2}\Gamma_{IV}^{ph}(t)$, respectively, similar to Eq.~(\ref{eq:solut_gelm}), which describe  the leakages of the populations of the OV and IV ionized states due to direct ionizations of the neutral Ne atom in the  Ne$^+(2\ell^{-1})$Ne to produce all possible final $\mbox{Ne}^+(2\ell^{-1})-\mbox{Ne}^+(2p^{-1})$   dicationic states.  Both $\Gamma_{OV}^{ph}(t)$ and $\Gamma_{IV}^{ph}(t)$ probabilities can be computed by   equations like Eq.~(\ref{eq:solut_gelm}). The time-independent imaginary term $-\frac{i}{2}\Gamma^{ICD}_{IV}(R)$ represents the losses of the population of the initial ICD state (i.e., IV-ionized state  Ne$^+(2s^{-1})$Ne) by ICD transition, i.e., the total ICD  rate. This rate  is given in the local approximation   by the Coulomb matrix element (\ref{eq:couplings_dec}) as 
\begin{equation}
\label{eq:tot_width}
\Gamma^{ICD}_{IV}(R)= 2\pi  \sum _j \vert V_{ICD}^{j}(R)\vert^2,
\end{equation} 
where the summation over  index $j$ runs over all possible ICD channels  [S5].

As was demonstrated in Ref.~[S6], the usual direct coupling $D_x(t)$ between the IV and OV ionized states through the laser field (by the $2s \leftrightarrow 2p$ excitation--deexcitation of Ne$^+$ in the present case) is augmented by an additional  time-dependent term $-\frac{i}{2} W(t)$.  This term appears only if the  photoionization from the OV-ionized  state and   ICD transition from IV ionized state  are simultaneously treated as required. Explicitly, it reads [S6]
\begin{equation}
\label{eq:solut_LIC}
W(t)=2\pi \sum_j \widetilde{d}^j_x(t) \left(V^{j} _{ICD} \right)^\dag   ,
\end{equation}
where the summation over   $j$ runs over all possible final dicationic states accessible by both channels. The  whole coupling between the `dressed' OV and IV ionized states is non-Hermitian and only operative as long as the pulse is on. The forth  row of the  Hamiltonian matrix (\ref{eq:fin2}) describes the nuclear dynamics on the  $\mbox{Ne}^+(2p^{-1})-\mbox{Ne}^+(2p^{-1})$  final OVOV-ionized state  produced by the emission of the ICD electron $\varepsilon_{ICD}$.  It is populated coherently  by the ICD transition from the IV-ionized state (matrix element (\ref{eq:couplings_dec})) and by  direct photoionization from the OV-ionized state (matrix element (\ref{eq:couplings_dir})), all at a given kinetic energy $\varepsilon_{ICD}$ of the emitted electron. The created  wave packet propagates on the `dressed'  potential energy surface  $V_{OVOV}(R)+ \varepsilon_{ph}+\varepsilon_{ICD}  -2\omega$.

The nuclear wave packet $  {\Psi}_{OVOV} (\varepsilon_{ph},\varepsilon_{ICD},t)   $   contains the information on the  coincident  photoelectron and ICD  electron spectrum. The coincident spectrum  can be computed as the norm of $  {\Psi}_{OVOV} (\varepsilon_{ph},\varepsilon_{ICD},t)   $  at  long times [S5]. Integration of the coincident spectrum  over the energy of the  photoelectron yields the individual ICD spectrum computed in the present study via
\begin{equation}
\label{eq:spectrum}
\sigma(\varepsilon_{ICD}) = \lim_{t\to\infty} \sum_j\int \langle  {\Psi}_{OVOV_j} (\varepsilon_{ph},\varepsilon_{ICD},t)  \vert {\Psi}_{OVOV_j} (\varepsilon_{ph},\varepsilon_{ICD},t) \rangle \, d\varepsilon_{ph}.
\end{equation} 
As a last  point we mention that at high intensities also the ionization of the initial ICD  states  Ne$^+(2s^{-1})$Ne to directly produce the $\mbox{Ne}^+(2s^{-1})-\mbox{Ne}^+(2p^{-1})$ states contributes to the final spectrum. To this end at least a third photon must be absorbed in the process~(\ref{eq:scheme}). In order to incorporate this mechanism in the calculations one has also to include the corresponding  electronic states in the ansatz (\ref{eq:anzatz}) which is straightforwardly  done.

As has been demonstrated in Refs.~[S7,S8], the `dressed'    states of a diatomic molecule can exhibit intersections  of the corresponding two-dimensional   complex potential energy surfaces  in the $R$ and $\theta$ space. The dynamical variables  $R$ and $\theta$ enter  the Hamiltonian matrix  (\ref{eq:fin2})  explicitly.  In the present case of ICD, the field couples the `dressed'  singly-ionized electronic states of  the Ne$^+_2$ (i.e., IV and OV states as discussed above). The rotational degree of freedom is involved in the nuclear dynamics only due to the presence of the laser field (the coupling matrix elements $H_{32}$ and $H_{23}$ are proportional to $\sin\theta$) and only when the pulse is on. Very important, the non-adiabatic couplings, i.e., matrix elements of the nuclear momenta along $R$ and $\theta$, between the two `dressed' electronic states are singular at these intersections [S17]. This gives rise to dramatic dynamical effects [S7,S8].

Without the  imaginary terms in the  Hamiltonian  (\ref{eq:fin2}), the two-dimensional potential energy surfaces $V_{OV}(R)$ and $V_{IV}(R)-\omega$ of the `dressed' electronic states   exhibit an intersection at  $\theta=0$ (where  $\sin\theta=0$), which is a conical intersection [S18-S20]. Due to the presence of  the ICD width and leakages by ionization in (\ref{eq:fin2}), the situation becomes more complicated. The two potential energy surfaces in $R$ and $\theta$  space  obtained by diagonalizing the electronic Hamiltonian  $\hat{\mathbf{H}}(R,\theta,t)- \hat{\textbf{T}}(R,\theta) $ in Eq.~(\ref{eq:fin2}) are now complex and generally exhibit two intersecting points at  which  the real as well as the imaginary parts of the two electronic energies become degenerate [S21]. This analogue of a conical  intersection in the continuum has been named   doubly intersecting complex energy surfaces (DICES). The impact of the \emph{light-induced} DICES on the nuclear dynamics accompanying resonant Auger decay of the core-excited CO molecule in intense laser fields has recently been studied in Ref.~[S8]. The non-adiabatic effects of DICES are naturally incorporated in the Hamiltonian matrix  (\ref{eq:fin2}) and, thus, were included in the present study of ICD.

To be able to carry out the two-dimensional calculations on the coupled complex energy surfaces we employed the efficient  Multi-Configuration Time-Dependent Hartree (MCTDH) method [S22] and code [S23]. The theoretical study of the nuclear dynamics problem requires the potential energies  of the electronic states participating in the process and electronic transition  rates between. In the calculations we have utilized the presently computed \emph{ab initio} energy curves (see Fig.~1 of the main text) and the \emph{ab initio} ICD transition rates reported in Ref.~[S24]. The values of the electron transition matrix elements were extracted from the experimental photoionization cross section of the Ne atom ($\sigma_{2p}={7.8}$~Mb at  28.4~eV [S25]) and the experimental $2s^{-1}\to 2p^{-1}$  radiative decay rate of Ne$^+$  ($\Gamma_r=4.8~\mu$eV   [S26]  corresponding to the radiative lifetime of $\tau_r\sim 0.14$~ns).

\makeatletter
\renewcommand*{\@biblabel}[1]{[S#1]}
\makeatother